\documentstyle[12pt]{article}

\def\nonequal{={\kern -1em}/ \,\,}
\def\pline{\,p\kern -0.47em /}
\def\bpline{\,{\bf P}\kern -0.6em\raise0.3ex \hbox{$/$}\,}
\def\qline{{q \kern-0.5em  / } }

\def\sm3{{\scriptscriptstyle (3)}}

\begin{document}


\begin{flushright}
FIAN/TD/00-15
\\
OHSTPY-HEP-T-00-020
\\
hep-th/0010248\\
\end{flushright}
\vspace{.5cm}


\begin{center}
{\Large \bf 
Light-cone approach to eleven dimensional 

\bigskip
supergravity}\footnote{
Talk given at International Conference on Quantization, Gauge Theory, and
Strings: Conference Dedicated to the Memory of Professor
Efim S. Fradkin, Moscow, Russia, 5-10 Jun 2000.} 

\vspace{1.5cm}

{R.R.Metsaev}


\vspace{1cm}

{\it
Department of Theoretical Physics, P.N. Lebedev Physical
Institute,\\ Leninsky prospect 53,  Moscow 117924, Russia

\vspace{0.5cm}

Department of Physics,
The Ohio State University  \\
Columbus, OH 43210-1106, USA\\
}


\vspace{1cm}

{\bf Abstract}
\end{center}

Manifestly supersymmetric formulation of eleven dimensional supergravity 
in the framework of light-cone approach is discussed.


\newpage

\section{Motivation and summary of results}

The long term motivation for our study of $11d$ supergravity is related to
conjectured iterrelation between superstring theory
and AdS higher spin massless field theory. Ten years ago 
E.S. Fradkin \cite{1}, based on studies made in \cite{p1}-\cite{p4}, 
put forward the idea that string theory and anti-de Sitter higher spin gauge 
theory,
though different, eventually may turn out to be 
different phases of one and the same unified field theory with new forces
mediated by higher spin gauge fields .
According to this conjecture string theory can be interpreteted as resulting
from some kind of spontaneous breakdown of higher spin symmetries.

To develop this idea it was conjectured recently \cite{met1} that 
{\it superstrings could be considered as living at the boundary
of $11$-dimensional AdS space while their unbroken (symmetric) phase is
realized as a theory of higher spin massless fields living in this
$AdS_{11}$ space}. Some discussion of this theme can be found in
\cite{met4} where it was demonstrated that if one restricts attention to
totally symmetric fields and make some mild assumption about the
(spontaneously) broken form of AdS theory Hamiltonian then  
{\it leading components of AdS massless totally symmetric arbitrary 
spin states become massive string states belonging to leading Regge
trajectory}.

As is well known the standard $11d$ supergravity \cite{crem} does not admit an
extension with a cosmological constant, i.e. does not have $AdS_{11}$ vacuum
\cite{des}(see also \cite{nic,sag}). One other hand,
in \cite{gun} certain massless $AdS_{11}$ graviton supermultiplet was
found\footnote{Related interesting discussion can be found in \cite{gm}.}.
This novel supermultiplet contains fields of the usual $11d$ supergravity plus
additional ones. One can expect that these additional fields may allow one 
to overcome no-go theorem and construct a consistent 
supergravity admitting $AdS_{11}$ ground state\footnote{
Certain massless $AdS_{11}$ graviton multiplet is also
predicted by eleven dimensional version of $AdS_{10}$ higher spin gauge 
theories discovered in \cite{vas10}.
These theories allow more or less straightforward generalization to $AdS_{11}$
\cite{vas11}. Since normally a tower of infinite higher spin fields 
contains of supergravity multiplet one expects that eleven dimensional
version of theories discussed in \cite{vas10} 
also describes some $AdS_{11}$ graviton multiplet.}.

The first step in this direction would be to find 
free field theoretic realization of this $AdS_{11}$ supermultiplet and
then try to construct interactions. Light-cone approach provides
self-contained setup to study these questions. The main advantage of light-cone
approach is that it allows one to discuss supersymmetric theories in terms of
unconstrained scalar superfields. Before attempting to study 
$AdS_{11}$ supergravity 
it would be interesting to consider the usual $11d$ supergravity which, to our
knowledge was not previously discussed in superfield light-cone gauge.

To discuss $11d$ supergravity we will exploit the method of \cite{dir} which
reduces the problem of finding a new  (light-cone gauge)  dynamical
system  to the problem of finding a new solution of commutation relations
of the  defining symmetry algebra  (in our case $11d$ Poincar\'e
superalgebra). In the past this method was  successfully applied for finding
manifestly supersymmetric formulations of various theories
\cite{GS5,BGS,GS6}\footnote{Derivation of light-cone formulation from
covariant Largangians of $N=4$ SYM was discussed in \cite{BLN,MAN}. 
Discussion of  higher spin massless fields may be found in
\cite{bbb}-\cite{met2}.}.
Despite many known examples of cubic vertices given in the literature,
constructing cubic vertices for concrete field theories is still a
challenging procedure. A general method essentially simplifying the
procedure of obtaining cubic interaction vertices was discovered in 
\cite{met20},
developed in \cite{met13,met16} and formulated finally  in \cite{met2}.
One of the characteristic features of this method is reducing
manifest transverse $so(d-2)$ invariance (which is $so(9)$ for $11d$
supergravity) to $so(d-4)$ invariance (which is $so(7)$ in this paper)\footnote{
Previously, reducing the manifest $so(d-2)$ symmetry to $so(d-4)$ was used 
to formulate 
superfield theory of $IIA$ superstrings \cite{GS6}. There this reducing was 
motivated by the desire to get unconstrained superfield formulation.
In \cite{met2} the main motivation for reducing was the desire to get the most 
general solution for cubic vertex for arbitrary spin fields of 
(super) Poincar\'e invariant theory. Discussion of $so(7)$ formalism in the
context of M(atrix) theory can be found in \cite{ple}.}. 
On the other hand, it is $so(7)$ symmetry that is manifest symmetry of 
unconstrained superfield formulation of $11d$ supergravity. In other words the
manifest symmetries of our method and the one of unconstrained superfield 
formulation of $11d$ supergravity match.
Here we demonstrate how the method of Ref. \cite{met2} 
works for the case of $11d$ supergravity.

Light-cone gauge $11d$ supergravity  can be formulated in
light-cone superspace which 
is based on position coordinates $x^\mu$,
and Grassmann position coordinates $\theta^\alpha$
\footnote{$\mu=0,1,\ldots 10$ are $so(10,1)$ vector indices,
$\alpha=1,\ldots, 8$ are 
$so(7)$ spinor index, $I,J,K=1,\ldots 9$ are  
$so(9)$ transverse indices, $i,j,k=1,\ldots,7$ are $so(7)$ transverse
indices.
Coordinates in light-cone directions are defined by 
$x^\pm \equiv (x^{10}\pm x^0)/\sqrt{2}$. Remaining transverse coordinates
$x^I$ are decomposed into $x^i$, $x^{R,L}$ where 
$x^{R,L}\equiv (x^8\pm {\rm i}x^9)/\sqrt{2}$. 
The scalar product of two  $so(9)$ vectors is decomposed then as
$X^I Y^I = X^i Y^i + X^RY^L + X^L Y^R$.
For momentum in light-cone
direction we use simplified notation
$\beta \equiv p^+$.}. 
In this light-cone superspace we introduce scalar
superfield $\Phi(x^\mu,\theta)$. Instead of position space it is 
convenient 
to use momentum space for all coordinates except the light-cone time  $x^+$.
This implies using $p^+$, $p^R$, $p^L$, $p^i$,
$\lambda^\alpha$, instead of $x^-$, $x^L$, $x^R$, $x^i$, $\theta^\alpha$
respectively.
Thus we consider the scalar superfield
$\Phi(x^+, p^+,p^R,p^L,p^i,\lambda)$ with the
following expansion in powers of Grassmann momentum $\lambda$

\begin{eqnarray}
\Phi(p,\lambda)&=&\beta^2 A + \beta\lambda^\alpha\psi^\alpha
+\beta\lambda^{\alpha_1}\lambda^{\alpha_2}A^{\alpha_1\alpha_2}
\nonumber\\
&+&\lambda^{\alpha_1}\lambda^{\alpha_2}\lambda^{\alpha_3}
\psi^{\alpha_1\alpha_2\alpha_3}
+\lambda^{\alpha_1}\ldots\lambda^{\alpha_4} A^{\alpha_1\ldots \alpha_4}
+\frac{1}{\beta}
(\epsilon\lambda^5)^{\alpha_1\alpha_2\alpha_3}
\psi^{\alpha_1\alpha_2\alpha_3*}
\nonumber\\
\label{supfield}
&-&\frac{1}{\beta}
(\epsilon\lambda^6)^{\alpha_1\alpha_2} A^{\alpha_1\alpha_2*}
-\frac{1}{\beta^2}(\epsilon\lambda^7)^\alpha\psi^{\alpha*}
+\frac{1}{\beta^2}(\epsilon\lambda^8) A^*\,,
\end{eqnarray}
where we use the notation\footnote{
In what follows a momentum $p$ as argument of the superfield $\Phi$ and
$\delta$--functions designates the set $\{p^I,\,\beta\}$. Also we do not
show explicitly the dependence of the superfield on evolution parameter 
$x^+$.}

\begin{equation}
(\epsilon\lambda^{8-n})^{\alpha_1\ldots \alpha_{n}}\equiv\frac{1}{(8-n)!}
\epsilon^{\alpha_1\ldots \alpha_{n}\alpha_{n+1}\ldots \alpha_8}
\lambda^{\alpha_{n+1}}\ldots\lambda^{\alpha_8}
\end{equation}
and $\epsilon^{\alpha_1\ldots \alpha_8}$ is the Levi-Civita tensor.
The only constraint which the superfield $\Phi$ should satisfy is the reality
constraint

\begin{equation}
\Phi(-p,\lambda)
=\beta^4\int d^8\lambda^\dagger e^{\lambda\lambda^\dagger/\beta}
(\Phi(p,\lambda))^\dagger\,.
\end{equation}
This constraint tells us that some fields in (\ref{supfield}) are related by
Hermitean conjugation.
In (\ref{supfield}) the component fields carrying even number of spinor 
indices describe bosonic fields

\begin{eqnarray}
\label{sup1}
&
A^{\alpha_1\ldots \alpha_4}(70) \sim \{h^{ij}(27^0),h^{RL}(1^0),
C^{ijk}(35^0),C^{RLi}(7^0)\}\,,
&
\\[7pt]
\label{sup2}
&
A^{\alpha_1\alpha_2}(28)
\sim  \{h^{Li}(7^{-1}),
C^{Lij}(21^{-1})\}\,,
\qquad
A=h^{LL}/\sqrt{2}\,,
&
\end{eqnarray}
while the fields with odd number of spinor indices are 
responsible for gravitino field.
Superscripts in (\ref{sup1}),(\ref{sup2}) indicate $J^{RL}$ charge.
Light-cone gauge action for the both free and interacting theory takes 
then the following standard form

\begin{equation}\label{lcact}
S=\int dx^+ \beta d\beta d^9pd^8\lambda
\Phi(-p,-\lambda){\rm i}\partial^-\Phi(p,\lambda)
+\int dx^+ P^-\,,
\end{equation}
where $\partial^- =\partial/\partial x^+$ and $P^-$ is Hamiltonian.
For free theory $P^-$ is given by the standard expression

\begin{equation}\label{fierep}
P_{(2)}^-=\int \beta d\beta d^9p d^8\lambda \Phi(-p,-\lambda)
\left(-\frac{p^Ip^I}{2\beta}\right) \Phi(p,\lambda)\,.
\end{equation}

Now let us discuss cubic interactions. General structure of 3 point
interaction vertices is obtainable from commutation relations  of
Poincar\'e superalgebra. Some of the latter lead  to the following 
expression for the Hamiltonian 

\begin{eqnarray}
\label{pm1}
P_{(3)}^-  = \int d\Gamma_3\prod_{a=1}^3 \Phi(p_a,\lambda_a)\, p_\sm3^-\,,
\end{eqnarray}
where 
the indices $a,b,c=1,2,3$ label three interacting superfields and

\begin{equation}\label{delfun}
d\Gamma_3\equiv \delta^{10}(\sum_{a=1}^3p_a)
\delta^8(\sum_{a=1}^3\lambda_a)
\prod_{a=1}^3 d\beta_a d^9p_a d^8\lambda_a\,.
\end{equation}
The Hamiltonian density $p_\sm3^-$ depends on
momenta $\beta_a$, transverse momenta $p_a^I$ and Grassmann momenta
$\lambda_a$.
The $\delta$- functions in (\ref{delfun}) respect
conservation laws for these momenta.

Next, using commutation relations of Hamiltonian with $J^{+I}$ and 
certain supercharges one finds 
that the Hamiltonian density $p_\sm3^-$, depends on
momenta $p_a^I$ and $\lambda_a$ in a special manner. Namely, it
turns out that $p_\sm3^-$ depends on $p_a^I$ and $\lambda_a$ through
the following quantities

\begin{equation}\label{pablab}
{\bf P}_{ab}^I\equiv p_a^I\beta_b-p_b^I\beta_a\,,
\qquad
\Lambda_{ab}\equiv \lambda_a\beta_b-\lambda_b\beta_a\,.
\end{equation}
The remarkable simplification is that
the new momenta ${\bf P}_{12}^I$, ${\bf P}_{23}^I$, ${\bf P}_{31}^I$ 
are not independent: all of them are
expressible  through ${\bf P}^I$ defined by\footnote{By using momentum
conservation laws for $p_a^I$ and $\beta_a$ it is easy to check that
${\bf P}_{12}^I={\bf P}_{23}^I={\bf P}_{31}^I={\bf P}^I$.}

\begin{equation}\label{defpi}
{\bf P}^I
=\frac{1}{3}\sum_{a=1}^3\check{\beta}_a p_a^I\,,
\qquad
\check{\beta}_a\equiv \beta_{a+1}-\beta_{a+2}\,,
\quad \beta_a\equiv \beta_{a+3}\,.
\end{equation}
The same happens for Grassmann momenta, i.e. due to momentum conservation
laws for $\beta_a$ and Grassmann momentum $\lambda_a$  
the new Grassmann momenta 
$\Lambda_{12}$, $\Lambda_{23}$, $\Lambda_{31}$ (see \ref{pablab})
are expressible in terms of one momentum $\Lambda$ defined by

\begin{equation}
\Lambda
=\frac{1}{3}\sum_{a=1}^3\check{\beta}_a \lambda_a\,.
\end{equation}
The usage of ${\bf P}^I$ and $\Lambda$ is advantageous since they are 
invariant under cyclic permutation of indices $1,2,3$ which label three
interacting fields.
Thus $p_\sm3^-$ 
is eventually the function of ${\bf P}^I$, $\Lambda$ and $\beta_a$,

\begin{equation}\label{p2v}
p_\sm3^- = p_\sm3^-({\bf P}, \Lambda,\beta_a)\,.
\end{equation}
The $p_\sm3^-$, by definition,  is a monomial of degree $k$ 
in ${\bf P}^I$. 
As is well known the original $11d$ supergravity is described by the vertex 
$p_\sm3^-$ involving terms of second order in transverse
momentum ${\bf P}^I$, i.e. we have to set $k=2$ . 
Let us for flexibility however keep $k$ to be 
arbitrary. Then the cubic vertex can be presented as

\begin{equation}
p_\sm3^- 
= {\bf P}^{I_1}\ldots {\bf P}^{I_k} p_\sm3^{-I_1\ldots I_k}
(\Lambda,\beta_a)\,.
\end{equation}
In general,
the $p_\sm3^{-I_1\ldots I_k}$ is a complicated $so(9)$ tensor
depending on Grassmann momentum $\Lambda$ and light-cone momenta $\beta_a$.
It is the finding this tensor that is the 
most difficult part of analysis of cubic
vertices. Note that $p_\sm3^-$ after reducing to $so(7)$ notation has the
decomposition

\begin{equation}
p_\sm3^- = ({\bf P}^L)^k p_\sm3^{-R\ldots R}
+({\bf P}^L)^{k-1}{\bf P}^i p_\sm3^{-iR\ldots R}
+\ldots 
+({\bf P}^R)^k p_\sm3^{-L\ldots L}\,.
\end{equation}
In \cite{met2} a method was suggested which allows one to express
$p_\sm3^{-I_1\ldots I_k}$,
which is $so(9)$ tensor, in terms of vertex $\tilde{V}_0$,
which is $so(7)$ scalar and has charge $k$ with respect to
$J^{RL}$. The general formula is 

\begin{equation}\label{p3v0}
p_\sm3^-({\bf P},\Lambda,\beta_a)
=({\bf P}^L)^k E_qE_\rho \tilde{V}_0(\Lambda,\beta_a)\,,
\end{equation}
where the operators $E_q$, $E_\rho$ are defined by relations

\begin{equation}\label{eq}
E_q=\exp(-q^j {\bf M}^{Lj}_\Lambda)\,,
\end{equation}

\begin{equation}\label{erho}
E_\rho\equiv \sum_{n=0}^k
(-\rho)^n
\frac{\Gamma(\frac{7}{2}+k-n)}{2^n n!\Gamma(\frac{7}{2}+k)}
({\bf M}_\Lambda^{Lj}{\bf M}_\Lambda^{Lj})^n\,,
\end{equation}
and we use the notation

\begin{equation}\label{newvar}
q^i\equiv
\frac{{\bf P}^i}{{\bf P}^L}\,,
\qquad
\rho\equiv
\frac{{\bf P}^i{\bf P}^i+2{\bf P}^R{\bf P}^L}
{2({\bf P}^L)^2}\,,
\qquad
\frac{{\bf P}^R}{{\bf P}^L}=
\rho-\frac{q^2}{2}\,.
\end{equation}
The vertex $\tilde{V}_0$
satisfies the following equations

\begin{eqnarray}
\label{mrltv0}
({\bf M}^{RL}_\Lambda-k)\tilde{V}_0=0\,,
\qquad
{\bf M}^{Ri}_\Lambda\tilde{V}_0=0\,,
\qquad
{\bf M}^{ij}_\Lambda\tilde{V}_0=0\,.
\end{eqnarray}
and it depends only on Grassmann momentum $\Lambda$ and
light-cone momenta $\beta_a$. The dependence on the transverse space 
momentum ${\bf P}^I$ is thus isolated explicitly.

The representation for $p_\sm3^-$ given in
Eqs.(\ref{p3v0})-(\ref{mrltv0}) is universal and  valid for 
arbitrary (super) 
Poincar\'e invariant theory. In order to get cubic vertices one 
needs (i) to find solutions to (\ref{mrltv0}); (ii) to 
insert $\tilde{V}_0$ and appropriate 
spin parts of angular momentum ${\bf M}^{IJ}$ fixed by representation
theory of super Poincar\'e algebra in (\ref{p3v0}). For the case under 
considerations the appropriate ${\bf M}^{IJ}$ are given by

\begin{equation}\label{MIJ}
{\bf M}^{RL}_\Lambda=\frac{1}{2}\theta_\Lambda\Lambda-2\,,
\quad
{\bf M}^{Ri}_\Lambda
=-\frac{1}{2\sqrt{2}}\hat{\beta}\theta_\Lambda\gamma^i\theta_\Lambda\,,
\quad
{\bf M}^{Li}_\Lambda
=\frac{1}{2\sqrt{2}\hat{\beta}}\Lambda\gamma^i\Lambda\,,
\end{equation}
where\footnote{$\gamma^i$ are usual $so(7)$ $\gamma$-matrices:
$\{\gamma^i,\gamma^j\}=2\delta^{ij}$. 
All of them are taken to be antisymmetric and hermitean.}

\begin{equation}
\hat{\beta}\equiv\beta_1\beta_2\beta_3\,,
\end{equation}
and the $\theta_\Lambda$ is defined by (anti)commutation relation 
$\{\theta_\Lambda, \Lambda\}=1$.

The remarkable property of the Eqs. (\ref{mrltv0})
is that normally they are quite simple to analyse\footnote{General solution to
these equations can be found in \cite{met2}.}. 
For instance, for the case of under consideration all what one needs is to 
analyse the first equation in (\ref{mrltv0}). Indeed, making use of 
expression for 
${\bf M}^{RL}$ given in (\ref{MIJ}) we find

\begin{equation}\label{dohl}
\Lambda\theta_\Lambda \tilde{V}_0=2(2-k)\tilde{V}_0\,.
\end{equation}
The operator $\Lambda\theta_\Lambda$  counts the degree of Grassmann
momentum $\Lambda$ involved in $\tilde{V}_0$ which, by
definition, cannot involve terms of negative power in $\Lambda$ i.e. 
eigenvalues of $\Lambda\theta_\Lambda$ must be non-negative. This
implies that vertices with terms higher than second order in  
${\bf P}^I$, i.e. when $k>2$, are forbidden. Note that 
it is the terms with $k=4$ and
$k=6$ that would correspond to supersymmetric extension of  higher
derivative terms like $R_{....}^2$ and $R_{....}^3$. Therefore, the fact
that vertices with $k=4$ and $k=6$ are forbidden implies that terms of
second and third order in Rieman tensor do not allow supersymmetric
extension\footnote{One important thing to note is that we proved absence 
of above
mentioned higher derivative terms  by using only commutation relations
between Hamiltonian $P^-$ and kinematical generators. Kinematical 
generators, by definition, are
the generators of super Poincar\'e algebra which have zero or positive $J^{+-}$
charge. It is
reasonable to think that kinematical generators do not  receive quantum
corrections. If this indeed would be the case then our result could be
considered as light-cone proof of nonrenormalization of $R_{....}^2$ and
$R_{....}^3$ terms in $11d$ supergravity. Note that we discuss
theory with 32 supercharges.
Study of $R_{....}^3$ terms in
string theory effective actions can be found in \cite{MTO}.
}. Thus the only value of $k$ allowed by 
Eq.(\ref{dohl}) is 
$k=2$, i.e. $\tilde{V}_0=const$\footnote{
Note that Eq.(\ref{dohl}) for $k=2$
tells us that $\tilde{V}_0$ does not depend on $\Lambda$ but then 
it still depends on light-cone momenta $\beta_a$. The fact that 
$\tilde{V}_0$ does not depend on $\beta_a$ too can be proved by using the
requirement that all (super)charge densities are polynomial in 
transverse momentum ${\bf P}^I$.},
and this leads to cubic vertex of the original $11d$ supergravity

\begin{equation}\label{p3v1}
p_\sm3^-({\bf P},\Lambda,\beta_a)
=\frac{\kappa}{3}({\bf P}^L)^2 E_qE_\rho\,,
\end{equation}
where $\kappa$ is the gravitational constant.
We choose normalization so that the cubic action for graviton field
obtainable from (\ref{lcact}),(\ref{p3v1}) coincides with the one of
the Einstein-Hilbert 

\begin{equation}\label{geact}
S_{EH}=\frac{1}{2\kappa^2}\int \sqrt{g}R\,,
\end{equation}
$R=R^{\mu\nu}{}_{\mu\nu}$,
$R^\mu{}_{\nu\lambda\sigma}=\partial_\lambda
\Gamma^\mu_{\nu\sigma}+\ldots$,
where we use the following expansion for metric tensor 
$g_{\mu\nu}=\delta_{\mu\nu}+\sqrt{2}\kappa h_{\mu\nu}$ and light-cone gauge
$h^{+\mu}=0$.

Making use of (\ref{p3v1}) and the formula 
for ${\bf M}^{Li}_\Lambda$ given
in (\ref{MIJ}) we can work out the 
explicit representation for cubic vertex in a
rather straightforward way

\begin{eqnarray}\label{exprep}
\frac{3}{\kappa} \, p_\sm3^-
&=&
{\bf P}^{L2}
-\frac{{\bf P}^L}{2\sqrt{2}\hat{\beta}}\Lambda\bpline\Lambda
+\frac{1}{16\hat{\beta}^2}(\Lambda\bpline\Lambda)^2
-\frac{{\bf P}_I^2}{9\cdot 16\hat{\beta}^2}(\Lambda\gamma^j\Lambda)^2
\nonumber\\
&+&\frac{{\bf P}^R}{9\cdot 16\sqrt{2}\hat{\beta}^3}
\Lambda\bpline\Lambda(\Lambda\gamma^j\Lambda)^2
+\frac{{\bf P}^{R2}}{2^7\cdot 63\hat{\beta}^4}
((\Lambda\gamma^i\Lambda)^2)^2\,,
\end{eqnarray}
where $\bpline \equiv {\bf P}^i\gamma^i$, 
${\bf P}_I^2\equiv {\bf P}^I{\bf P}^I$.
Thus we have two equivalent representations for $11d$ supergravity 
cubic interaction vertex given by (\ref{p3v1}) and (\ref{exprep}).
The representation (\ref{exprep}) being manifest in 
Grassmann momentum $\Lambda$ is not convenient, however, in
calculations. In contrast, the representation (\ref{p3v1}) 
does not show explicitly the dependence on $\Lambda$.
However the remarkable feature of representation (\ref{p3v1}) is that it 
is expressed entirely in terms of spin
operator ${\bf M}^{Li}$ which has clear algebraic properties.  
For this reason it is the representation (\ref{p3v1}) that is the 
most convenient in 
calculations. As compared to 
(\ref{exprep}), the representation (\ref{p3v1}) is
universal. For instance the cubic vertices of $IIA$ SURGA and $N=1$ 
ten-dimensional SYM have similar form.

\section{Conclusion}

We have discussed  the light-cone gauge formulation of 
usual $11d$ supergravity. The formulation is given entirely in terms of
light-cone scalar superfield allowing us to treat all component fields on
an equal footing.
Because the formalism we presented is algebraic in nature it can be
extended to AdS spacetime in a relative straightforward way.
Comparison of  this formalism with other approaches   available in the
literature leads us to  the conclusion
that this is a very efficient formalism.

The formulation presented here should have a number of interesting
applications and generalizations, some of which are:

(i) generalization to $AdS_{11}$ spacetime and study 
of massless $AdS_{11}$ graviton supermultiplet 
found in \cite{gun}.

(ii) application of
manifestly supersymmetric light-cone formalism to the study 
of the various aspects of M-theory along the lines 
\cite{ple2}--\cite{van};

(iii) generalization to cubic vertices of type IIB
supergravity in $AdS_5\times S^5$ background \cite{MRR}
and then to strings in this background \cite{MT,KRR,MTT}.

\bigskip
\bigskip

\noindent{\bf Acknowledgements.} 
This work was supported in part by
the DOE grant DE-FG02-91ER-40690, by the INTAS project 991590,
and by the RFBR Grant No.99-02-17916.


\newpage

\begin{thebibliography}{30}


\bibitem{1}
E.S.Fradkin
`Higher spin symmetries and the problem of
unification of all Interactions' 1992, In lecture given at ETH, Preprint
ETH-TH/92-10)


\bibitem{p1}
E.~S.~Fradkin and M.~A.~Vasiliev,
Annals Phys.\  {\bf 177}, 63 (1987).


\bibitem{p2}
E.~S.~Fradkin and M.~A.~Vasiliev,
Phys.\ Lett.\  {\bf B189}, 89 (1987).


\bibitem{p3}
E.~S.~Fradkin and R.~R.~Metsaev,
Class.\ Quant.\ Grav.\  {\bf 8}, L89 (1991).


\bibitem{p4}
E.~S.~Fradkin and V.~Y.~Linetsky,
Nucl.\ Phys.\  {\bf B350}, 274 (1991).

\bibitem{met1}
R.~R.~Metsaev,
Nucl.\ Phys.\  {\bf B563}, 295 (1999)
[hep-th/9906217].


\bibitem{met4}
R.R.~Metsaev,
``IIB supergravity and various aspects of light-cone formalism in
AdS  space-time,''
Talk given at International Workshop on Supersymmetries 
and Quantum Symmetries
(SQS 99), Moscow, Russia, 27-31 Jul 1999;
hep-th/0002008.


\bibitem{crem}
E.~Cremmer, B.~Julia and J.~Scherk,
Phys.\ Lett.\  {\bf B76}, 409 (1978).


\bibitem{des}
K.~Bautier, S.~Deser, M.~Henneaux and D.~Seminara,
Phys.\ Lett.\  {\bf B406}, 49 (1997)
[hep-th/9704131];
S.~Deser,
``D = 11 supergravity revisited,''
hep-th/9805205.

\bibitem{nic}
H.~Nicolai, P.~K.~Townsend and P.~van Nieuwenhuizen,
Lett.\ Nuovo Cim.\  {\bf 30}, 315 (1981).


\bibitem{sag}
A.~Sagnotti and T.~N.~Tomaras,
``Properties Of Eleven-Dimensional Supergravity,''
CALT-68-885.


\bibitem{gun}
M.~Gunaydin,
Nucl.\ Phys.\  {\bf B528}, 432 (1998)
[hep-th/9803138].


\bibitem{gm}
M.~Gunaydin and D.~Minic,
Nucl.\ Phys.\  {\bf B523}, 145 (1998)
[hep-th/9802047].

\bibitem{vas10}
M.~A.~Vasiliev,
Phys.\ Lett.\  {\bf B257}, 111 (1991).

\bibitem{vas11}
M.~A.~Vasiliev, private communication (Nov. 1999)


\bibitem{dir}
P.~A.~Dirac,
Rev.\ Mod.\ Phys.\  {\bf 21}, 392 (1949).


\bibitem{GS5}
M.~B.~Green and J.~H.~Schwarz,
Phys.\ Lett.\  {\bf B122}, 143 (1983).

\bibitem{BGS}
L.~Brink, M.~B.~Green and J.~H.~Schwarz,
Nucl.\ Phys.\  {\bf B223}, 125 (1983).


\bibitem{GS6}
M.~B.~Green and J.~H.~Schwarz,
Nucl.\ Phys.\  {\bf B243}, 475 (1984).


\bibitem{BLN}
L.~Brink, O.~Lindgren and B.~E.~Nilsson,
Nucl.\ Phys.\  {\bf B212}, 401 (1983).

\bibitem{MAN}
S.~Mandelstam,
Nucl.\ Phys.\  {\bf B213}, 149 (1983).


\bibitem{bbb}
A.~K.~Bengtsson, I.~Bengtsson and L.~Brink,
Nucl.\ Phys.\  {\bf B227}, 31 (1983).
Nucl.\ Phys.\  {\bf B227}, 41 (1983).


\bibitem{mold}
R.~R.~Metsaev,
Mod.\ Phys.\ Lett.\  {\bf A6}, 359 (1991).
Mod.\ Phys.\ Lett.\  {\bf A8}, 2413 (1993).


\bibitem{met20}
R.~R.~Metsaev,
Class.\ Quant.\ Grav.\  {\bf 10}, L39 (1993).


\bibitem{met13}
R.~R.~Metsaev,
Phys.\ Lett.\  {\bf B309}, 39 (1993).


\bibitem{met16}
E.~S.~Fradkin and R.~R.~Metsaev,
Phys.\ Rev.\  {\bf D52}, 4660 (1995).


\bibitem{met2}
R.~R.~Metsaev,
``Cubic interaction vertices for higher spin fields,''
Talk given at 2nd International Sakharov Conference on 
Physics, Moscow, Russia,
20-23 May 1996. 
hep-th/9705048.


\bibitem{ple}
J.~Plefka and A.~Waldron,
Nucl.\ Phys.\  {\bf B512}, 460 (1998)
[hep-th/ 9710104].


\bibitem{MTO}
R.~R.~Metsaev and A.~A.~Tseytlin,
Phys.\ Lett.\  {\bf B185}, 52 (1987).




\bibitem{ple2}
A.~Dasgupta, H.~Nicolai and J.~Plefka,
JHEP {\bf 0005}, 007 (2000)
[hep-th/0003280].


\bibitem{ple3}
R.~Helling, J.~Plefka, M.~Serone and A.~Waldron,
Nucl.\ Phys.\  {\bf B559}, 184 (1999)
[hep-th/9905183].


\bibitem{gr}
T.~Dasgupta, M.~R.~Gaberdiel and M.~B.~Green,
JHEP {\bf 0008}, 004 (2000)
[hep-th/0005211].


\bibitem{van}
M.~B.~Green, H.~Kwon and P.~Vanhove,
Phys.\ Rev.\  {\bf D61}, 104010 (2000)
[hep-th/9910055].

\bibitem{MRR}
R.~R.~Metsaev,
Phys.\ Lett.\  {\bf B468}, 65 (1999)
[hep-th/9908114].


\bibitem{MT}
R.~R.~Metsaev and A.~A.~Tseytlin,
Nucl.\ Phys.\  {\bf B533}, 109 (1998)
[hep-th/9805028];
Phys.\ Lett.\  {\bf B436}, 281 (1998)
[hep-th/9806095];
``Superstring action in AdS(5) x S**5: kappa-symmetry light cone gauge,''
hep-th/0007036.


\bibitem{KRR}
R.~Kallosh, J.~Rahmfeld and A.~Rajaraman,
JHEP {\bf 9809}, 002 (1998)
[hep-th/9805217].



\bibitem{MTT}
R.~R.~Metsaev, C.~B.~Thorn and A.~A.~Tseytlin,
``Light-cone superstring in AdS space-time,''
hep-th/0009171.


\end{thebibliography}
\end{document}